\documentstyle[12pt]{article}
\newcommand{\beq}{\begin{equation}}
\newcommand{\eeq}{\end{equation}}
\newcommand{\bea}{\begin{eqnarray}}
\newcommand{\eea}{\end{eqnarray}}

\newcommand{\g}{\gamma}
\newcommand{\k}{\kappa}

\newcommand{\et}{{\tilde e}}

\newcommand{\wt}{{\tilde \omega}}

\newcommand{\thb}{{\overline \theta}}

\newcommand{\p}{\phi}
\newcommand{\w}{\omega}

\renewcommand{\d}{\delta}
\renewcommand{\l}{\lambda}
\renewcommand{\L}{\Lambda}
\renewcommand{\b}{\beta}
\renewcommand{\a}{\alpha}
\newcommand{\n}{\nu}
\newcommand{\m}{\mu}
\newcommand{\r}{\rho}
\newcommand{\s}{\sigma}
\newcommand{\th}{\theta}

\newcommand{\oh}{\frac{1}{2}}
\newcommand{\oq}{\frac{1}{4}}

\newcommand{\non}{\nonumber}
\newcommand{\zp}{\frac{d^Dp}{(2\pi)^D}}

\newcommand{\rf}[1]{(\ref{#1})}
\newcommand{\ra}{\rightarrow}
\newcommand{\beaa}{\begin{eqnarray}{c}}
\newcommand{\eeaa}{\end{eqnarray}{c}}
\newcommand{\gpb}{\Gamma(-D/2, \overline x_{B+})}
\newcommand{\gmb}{\Gamma(-D/2, \overline x_{B-})}
\newcommand{\gpf}{\Gamma(-D/2, \overline x_{F+})}
\newcommand{\gmf}{\Gamma(-D/2, \overline x_{F-})}
\newcommand{\xpb}{x_{B+}}
\newcommand{\xmb}{x_{B-}}
\newcommand{\xpf}{x_{F+}}
\newcommand{\xmf}{x_{F-}}
\newcommand{\Lover}{{\Lambda^D\over 2(4\pi)^{D/2}}}
\newcommand{\dnbf}{\Delta (n)_{BF}}
\newcommand{\bold}{\load{\normalsize}{\bf}}
\newcommand{\dx}{\Delta (x)_{BF}}
\newcommand{\dxdue}{\Delta (x^2)_{BF}}
\newcommand{\dxtre}{\Delta (x^3)_{BF}}
\newcommand{\dxlog}{\Delta (x\ln{x})_{BF}}
\newcommand{\dxduelog}{\Delta (x^2\ln{x})_{BF}}
\newcommand{\dxtrelog}{\Delta (x^3\ln{x})_{BF}}
\newcommand{\ox}{O[\Delta (x^4\ln{x})_{BF}]}

\newcommand{\cdp}{\cos^2\left({D\pi\over 2}\right)}
\newcommand{\sdp}{\sin^2\left({D\pi\over 2}\right)}
\newcommand{\D}{\Delta}
\begin{document}

\addtolength{\baselineskip}{0.20\baselineskip}
\hfill NBI-HE-93-37

\hfill gr-qc/9308012

\hfill August 1993
\begin{center}

\vspace{36pt}
{\large \bf WHY IS SPACETIME LORENTZIAN?}
\footnote{Supported by the U.S. Department of Energy under Grant
No. DE-FG03-92ER40711.}

\end{center}

\vspace{24pt}

\begin{center}
{\sl A. Carlini
\footnote{Supported by an INFN grant at Lawrence Berkeley
Lab., 1 Cyclotron Rd., Berkeley CA 94720}
\footnote{Email: carlini@theorm.lbl.gov}
and J. Greensite}
\footnote{Permanent address: Physics and Astronomy Dept., San Francisco
State University, San Francisco CA 94132.}
\footnote{Email: greensit@stars.sfsu.edu, greensite@nbivax.nbi.dk} \\

\vspace{12pt}

The Niels Bohr Institute \\
Blegdamsvej 17 \\
DK-2100 Copenhagen \O ~~Denmark \\

\vfill
{\bf Abstract}
\vspace{12pt}
\end{center}

    We expand on the idea that spacetime signature should be treated
as a dynamical degree of freedom in quantum field theory. It has been
argued that the probability distribution for signature, induced by
massless free fields, is peaked at the Lorentzian value uniquely
in D=4 dimensions.  This argument is reviewed, and certain consistency
constraints on the generalized signature (i.e. the tangent space metric
$\eta_{ab}(x)=\mbox{diag}[e^{i\theta(x)},1,1,1]$) are derived.
It is shown that only one dynamical "Wick angle" $\theta(x)$ can
be introduced in the generalized signature, and the magnitude of
fluctuations away from Lorentzian signature $\delta \theta = \pi -
\theta$ is estimated to be of order $(l_P/R)^3$, where $l_P$ is the
Planck length, and $R$ is the length scale of the Universe.
For massless fields, the case of D=2 dimensions and the case of
supersymmetry are degenerate, in the sense that no signature is
preferred.  Mass effects lift this degeneracy, and we show that
a dynamical origin of Lorentzian signature is also possible for
(broken) supersymmetry theories in D=6 dimensions, in addition to
the more general non-supersymmetric case in D=4 dimensions.

\vfill

\newpage

\section{Introduction}

   A theorem of matrix algebra states that any real symmetric matrix
$M$ can be written in the form $M=SDS^T$, where $S$ is a real-valued
matrix and $D$ is a diagonal matrix with values $\pm 1$ and $0$ along the
diagonal. These diagonal entries are known as the "signature" of the
matrix $M$, and are unique up to permutations.  The metric of general
relativity is normally taken to be
a real symmetric matrix, and can therefore be written in the form
$g_{\m\n} = e^a_\m \eta_{ab} e^b_\n$, where $\eta_{ab}$ is the diagonal
tangent-space metric.  It has been known since the work of Minkowski that
physical spacetime has a Lorentzian signature
$\eta = \mbox{diag}[-1,1,1,1]$.

    The Einstein field equations $G_{\m\n}=-\k T_{\m\n}$ do not, however,
impose any particular restriction on spacetime signature; in fact, they
do not refer to signature at all.  There is nothing inherent in
classical general relativity which either fixes the spacetime signature to
be Lorentzian, or even, given that the signature is initially
Lorentzian, forces spacetime in all cases to remain Lorentzian.
In this connection, several authors \cite{Gibbons,Ellis,Dereli} have
constructed solutions to the Einstein equations which
evolve from Euclidean to Lorentzian signature.
If signature-changing processes can occur classically, then they can presumably
also occur quantum-mechanically (in fact, such speculations are
not uncommon in quantum cosmology, see e.g.
\cite{Gibbons,Sakharov,Vilenkin}).  This then raises
the question of why it is, if other signatures are dynamically accessible,
that spacetime is found to be everywhere Lorentzian.

   An explanation of the origin of Lorentzian signature at the quantum
level could take several forms.  The simplest, and in our opinion the
least satisfying, is to simply assume the existence of a constraint such
as $\det(g)<0$ in the functional integration measure (this can also be
done in tetrad formulation by imposing a fixed $\eta_{ab}$).  Another
possibility is that for some reason (perhaps the absence of certain anomalies),
Lorentzian signature is the only consistent choice at the quantum
level, as may be the case in string theory \cite{Tata}.
Finally, there could be dynamical reasons why Lorentzian signature is
preferred over other signatures.

   In a recent article \cite{Me} one of us suggested a dynamical origin
for Lorentzian signature; the idea is to generalize the concept
of Wick rotation in path-integral quantization.  Rather than viewing
Wick rotation as a technicality necessary for convergence of the
path-integral, the Wick angle $\th$ is treated as a dynamical degree
of freedom, which is free to fluctuate.  The tangent space metric
then has the form

\beq
    \eta = \mbox{diag}[\exp(i\th),1,...,1]
\eeq
In ref. \cite{Me} the one-loop
(complex-valued) effective potential $V(\th)$, generated by massless fields,
was calculated.  It was found
that if the number of fermionic degrees of freedom
exceeds the number of bosonic degrees of
freedom, then $Re[V]$ has a minimum and $Im[V]$ is stationary, uniquely
in $D=4$ dimensions, at $\th=\pm \pi$, corresponding to Lorentzian signature.
In this way a relation was found between the dimension of spacetime,
the signature of spacetime, and the presence of the factor of $i$ in
the path amplitude $\exp[iS]$.

    The present article expands further on the idea of dynamical
signature. The results of ref. \cite{Me}, in a flat background space,
are reviewed in section 2, and a quantum evolution equation in
non-Lorentzian spacetime is proposed. Consistency conditions in curved
spacetime are discussed in section 3.  On the grounds that (i) the
tangent space metric $\eta_{\m\n}$ is flat; (ii) the number of gravitational
degrees of freedom is independent of $\eta_{\m\n}$; and (iii) a spin
connection with appropriate properties is obtained in the Dirac action,
certain strong constraints on the functional dependence of the Wick angle
are deduced.  These constraints turn out to be crucial in suppressing what
would otherwise be unacceptably large quantum fluctuations away from
Lorentzian signature.  It is also shown that it is only possible to have
a single dynamical Wick angle satisfying the constraints; a tangent-space
metric with multiple angles

\beq
    \eta = \mbox{diag}[\exp(i\th_1),\exp(i\th_2)...,\exp(i\th_D)]
\eeq
is ruled out.  In section 4 it is shown that the cosmological
constant at one-loop cannot be subtracted by a counterterm for all values of
$\th$; in fact, if the Wick angle is dynamical, the cancellation can only be
made in D=4 dimensions at $\th=\pm \pi$.
In section 5 we extend the results of ref. \cite{Me} by including
mass terms for the fermionic and bosonic fields.
Again requiring a minimum/stationarity condition for the one-loop
effective potential $V(\th)$ we show that, in addition to the case
of D=4 dimensions found previously, there is also a possible solution
for (broken) supersymmetric theories, at $\th=\pm \pi$ and D=6.
Section 6 contains the conclusions.

\section{The Dynamical Wick Angle}

  In the path-integral formulation of quantum field theory,
it is required to evaluate Feynman path-integrals of the form

\beq
       Z_F = \int d\m(e,\p,\psi,\overline{\psi}) \;
\exp\left[-i \int d^Dx \sqrt{-g} \cal{L}
\right]
\label{ZF}
\eeq
where $d\m(e,\p,\psi,\overline{\psi})$ is the integration measure for
the tetrads, and other bosonic ($\p$) and fermionic ($\psi,\overline{\psi}$)
fields.  The restriction to Lorentzian spacetime is enforced by
working with a fixed signature

\bea
            g_{\m\n} &=& e^a_\m \eta_{ab} e^b_\n
\non \\
            \eta_{ab} &=& \mbox{diag}[-1,1,...,1]
\eea
and in the case of a flat background, one simply sets $g_{\m\n}=\eta_{\m\n}$.
However, in order to define propagators and other correllators,
it is necessary to improve the convergence properties of the
Feynman amplitude $e^{iS}$. Note that even a zero-dimensional gaussian
integral

\beq
      \int^{\infty}_{-\infty} dx \; x^{2n} e^{ix^2}
\eeq
does not converge, when evaluated numerically, for $n \ge 1$.  Convergence
can be improved by either adding a small imaginary mass term (the $i\epsilon$
prescription), or else by rotating the time axis into the complex plane.
Rotating $t \ra it$ gives the Euclidean path-integral

\beq
       Z_E = \int d\m(e,\p,\psi,\overline{\psi}) \;
\exp \left[- \int d^Dx \sqrt{g} \cal{L} \right]
\label{ZE}
\eeq
where this time

\beq
        \eta_{ab} = \mbox{diag}[1,1,...,1]
\label{Euclid}
\eeq
Comparing the Feynman and Euclidean path-integrals, it is easy to
write down a path-integral which interpolates between them, namely

\beq
      Z = \int d\m(e,\p,\psi,\overline{\psi}) \;
\exp\left[- \int d^Dx \sqrt{g} \cal{L}
\right]
\label{ZG}
\eeq
where
\beq
        \eta_{ab} = \mbox{diag}[e^{i\th},1,...,1]
\label{general}
\eeq
The Euclidean theory is obtained for $\th=0$ and the Feynman theory
for $\th=\pi$, with the correct $i \epsilon$ prescription for propagators
automatically supplied as $\th \ra \pi$.

   Motivated by Lorentzian to Euclidean signature change
at the classical level, we now consider the possibility that the
"signature" of eq. \rf{general} is free to fluctuate; i.e. that $\th$
is a dynamical degree of freedom.\footnote{We will continue to refer to
the (complex) entries of $\eta$ as the "signature", although this is
admittedly an abuse of the mathematical terminology.}  This requires,
of course, some generalization of quantum mechanics.  Consider a fixed Wick
angle $\th$ anywhere in the range $-\pi<\th<\pi$ (note that $|\th|>\pi$
is ruled out because the kinetic term in the bosonic field
action would be unbounded from below).  Assuming a flat-space
($e^a_\m=\d^a_\m$) background and denoting the fields collectively by
$\phi$, the path-integral definition of transition amplitudes is

\beq
  G[\phi_{f},t_f|\phi_{i},t_i] \equiv  \int^{\phi_f}_{\phi_i} d\m(\p) \;
\exp\left[- \int^{t_f}_{t_i}dt \int d^{D-1}x \sqrt{g} \cal{L}
\right]
\eeq
and we obtain, by the usual arguments, the generalized Schrodinger equation

\beq
       \partial_t \Psi[\p] = -e^{i\th/2} H \Psi[\p]
\label{Seq}
\eeq
where $H$ is the standard (and hermitian) Hamiltonian.  For any
$\th \ne \pm \pi$ the norm of $\Psi$ can change.  Therefore, to conserve
probability, $\Psi$ must be interpreted as supplying {\it relative}
probabilities, or, equivalently,

\beq
       <Q> \equiv {<\Psi|Q|\Psi> \over <\Psi|\Psi>}
\label{exp}
\eeq
Equations \rf{Seq} and \rf{exp} together give

\bea
      \partial_t <Q> &=& \sin{\th \over 2}<i[H,Q]>
\non \\
&-& \cos{\th \over 2} \{<HQ+QH> - 2<Q><H>\}
\label{evolve}
\eea
Providing $Q$ is hermitian, this evolution equation preserves the reality
of observables, and satisfies conservation of probability.

  On the other hand, for $\th \ne \pm \pi$, conservation of energy is
violated

\beq
      \partial_t <H> = - 2 \cos{\th \over 2}<(H-<H>)^2>
\eeq
(along with Lorentz invariance), and an arbitrary initial state $\Psi_{in}$
will eventually relax either to the ground state $\Psi_0$, or else to the
lowest energy eigenstate $\Psi_E$ for which $<\Psi_{in}|\Psi_E>\ne 0$.
There are, of course, very stringent observational limits on non-conservation
of energy; see, e.g. ref. \cite{energy}.  The first problem, for a theory
in which the Wick angle is allowed to fluctuate, is to show that the
probability distribution is peaked at Lorentzian signature $\th=\pi$.  The
next problem is to show that fluctuations away from Lorentzian signature are
so strongly suppressed that observational bounds on energy conservation are
not violated.

    To study the first problem, we need to compute the effective
potential $V_{eff}(\th)$
for the Wick angle, which is generated after integrating
out all other fields.  In ref. \cite{Me} this was computed for massless fields
at one-loop level.  The calculation requires some assumptions about the
$\th$-dependence of the integration measure, which is otherwise just taken
proportional to the (real-valued) DeWitt measure.  The following assumptions
were made:

\begin{description}
\item[~~1.] ~For free fields of mass $m$, the contributions to
$Z$ in eq. \rf{ZG} from
each (propagating) bosonic degree of freedom are equal, and inverse to
the contributions from each fermionic degree of freedom.  Thus, e.g.,
$Z=1$ at any $\th$ for a supersymmetric combination of free fields.
\item[~~2.] ~The integration measure for scalar fields is given by the
real-valued, invariant volume measure (DeWitt measure) in superspace
$d\m(\p) = D\p \sqrt{|G|}$, where $G$ is the determinant of the
scalar field supermetric $G(x,y)=\sqrt{g}\d(x-y)$.
\end{description}

   Under these assumptions, the one-loop contribution to $V_{eff}(\th)$ due
to a massless scalar field propagating in flat ($g_{\m\n}=\eta_{\m\n}$)
space is

\beq
      \exp[-\int d^Dx V_S(\th)] = det^{-\oh}[- \sqrt{\eta}\eta^{ab}
\partial_a \partial_b]
\label{det}
\eeq
and heat-kernel regulation of the determinant gives

\bea
    V_S(\th) &=& -\oh \int^{\infty}_{1/\L^2} {ds \over s}\int \zp
\; \exp[- s(e^{-i\th/2}p_0^2 + e^{i\th/2}\vec{p}^{\;2})]
\non \\
       &=& -{\L^D \over D(4\pi)^{D/2}} \exp[-i(D-2)\th/4]
\label{proper}
\eea
where $\L$ is a high-momentum cutoff which, given the
non-renormalizability of gravity, is taken to exist at the Planck
scale.  For our purposes, the choice of heat-kernel regularization
is essentially unique.  Zeta function and dimensional regulation
methods contain implicit subtractions which remove non-logarithmically
(e.g. quadratically) divergent terms; these happen to be the terms of
interest here. On the other hand, a naive momentum-space cutoff does
not uniformly respect the spacetime symmetries at $\th=0,\pm \pi$.
A cutoff such as $k_0^2 + \vec{k}^2 < \L^2$, which is appropriate for
the Euclidean case at $\th=0$, is clearly asymmetric at $\th=\pm \pi$,
and the reverse is true for, say, $|k_0^2 - \vec{k}^2| < \L^2$.
The same objection applies to a lattice cutoff; moreover, a regular
lattice, even at $\th=0$, does not respect the full $O(D)$ symmetry.
We are looking for a regulator which respects the symmetries at
$\th=0,\pm \pi$, and which interpolates smoothly in the range
$\th \in [-\pi,\pi]$.  With these requirements, the choice of
heat-kernel regularization seems almost unavoidable.
In connection with the assumptions about the measure,
it is worth noting that these lead, for any spin, to a
contribution which can be regulated at all
$\th \in [-\pi,\pi]$ by the heat kernel technique.

   For $n_B$ massless, propagating, bosonic degrees of freedom,
and $n_F$ massless fermionic degrees of freedom, the one-loop
contribution to $V_{eff}(\th)$ becomes

\beq
    V(\th) =  (n_F-n_B){\L^D \over D(4\pi)^{D/2}} \exp[-i(D-2)\th/4]
\label{V}
\eeq
This potential is complex.  We therefore look for a value of $\th$
in the range $[-\pi,\pi]$ for which, simultaneously, (i) $Re(V)$ is a
minimum; and (ii) $Im(V)$ is stationary.  These conditions together
give us

\bea
    \left. \begin{array}{rr}
          \cos{[(D-2)\th/4]} = 0 \\
                                     \\
         \min{[Re[V(\th)]]}= 0     \\
  \end{array} \right\} \mbox{~~~$\th \in [-\pi,\pi]$}
\label{conditions}
\eea
In searching for a solution of \rf{conditions}, there are
five cases to consider:

\begin{description}
\item{~~I.} ~$n_F<n_B$.  Then $\mbox{minRe}$[$V]<0 \ra$ no solution.
\item{~II.} ~$n_F=n_B$ or $D=2$.  Then $V(\th)$ is independent of
$\th$, and no $\th$ is preferred.
\item{III.} ~$n_F>n_B$ and ${D-2 \over 4}\pi < {\pi \over 2}$.
Then $\mbox{minRe}$[$V]>0 \ra$ no solution.
\item{~IV.} ~$n_F>n_B$ and ${D-2 \over 4}\pi > {\pi \over 2}$.
Then $\mbox{minRe}$[$V]<0 \ra$ no solution.
\item{~~V.} ~$n_F>n_B$ and ${D-2 \over 4}\pi = {\pi \over 2}$.  In
this case, both conditions are satisfied at $\th = \pm \pi$,
which corresponds to Lorentzian signature.  The equality
${D-2 \over 4}\pi = {\pi \over 2}$ can, of course, only be achieved
for a spacetime dimensionality $D=4$.
\end{description}

   Since case (V), above, is the unique solution of the conditions
\rf{conditions}, we have found an interesting connection between
spacetime signature and spacetime dimension: Lorentzian signature
seems to be singled out by the dynamics only in $D=4$ dimensions.

   It is natural to look for generalizations.  For example, just as
the $\eta$ of \rf{general} interpolates between Euclidean and Lorentzian
signature, one might consider metrics interpolating between a Lorentzian
and a "two-time" signature, i.e.

\beq
   \eta_{ab} = \mbox{diag}[-1,e^{i\th},1,...,1]
\eeq
However, it is easy to see that the kinetic term of a functional
integral with such a signature is, for general $\th \ne \pm \pi$,
unbounded from below.  One the other hand, one could instead
consider tangent space metrics with two or more dynamical
"Wick angles", e.g.
\beq
   \eta_{ab} = \mbox{diag}[e^{i\th_1},e^{i\th_2},...,e^{i\th_D}]
\label{multiple}
\eeq
with the $\{\th_n\}$ suitably restricted to ensure the boundedness
of the action.  Finally, it is important to investigate the expected
size of fluctuations away from Lorentzian signature.  The magnitude
of such fluctuations, which violate both Lorentz invariance and
energy conservation, would have to be extremely small to be consistent
with experiment.  These issues will be discussed in the next section.

\section{The Wick Angle in Curved Spacetime}

   $V(\th)$ was computed above for constant $\th$. Before tackling
the question of fluctuations away from $\th=\pi$,
we should ask whether there are any
restrictions that should be imposed on the functional dependence of
$\th(x)$, apart from the condition that $|\th|<\pi$.  Since $\eta_{\m\n}$
is supposed to be a generalization of the flat space metric, it is
reasonable to impose the condition that the Riemann tensor computed
from $g_{\m\n}=\eta_{\m\n}$ vanishes.  To put it another way,
signature should not, by itself, generature
curvature.\footnote{Otherwise we would really be dealing with a fully
complex general relativity, and we should consider complex general coordinate
transformations, resulting in complex coordinates.}
In Cartesian coordinates,
with $e^a_\m=\d^a_\m$, this means that $\th$ depends only on the time
coordinate $\th=\th(t)$.  The obvious generalization of $\th=\th(t)$
to curved spacetime is

\beq
       \partial_\m \th = f e^0_\m \propto e^0_\m
\eeq
This leads to a consistency condition
\beq
       0 = (\partial_\m \partial_\n - \partial_\n \partial_\m) \th
= \partial_\m (fe^0_\n) - \partial_\n (fe^0_\m)
\label{con1}
\eeq
which imposes some extra constraints on $e^0_\m$.  In fact, \rf{con1}
is satisfied by

\bea
          \th &=& \th(T(x))
\non \\
           e^0_\m &=& \partial_\m T(x)
\label{con2}
\eea
It will now be shown that these conditions on $e^0_\m$ and $\th$ are
required by two other, quite different arguments, one of which
concerns the number of degrees of freedom of the gravitational field.

   In $D$ dimensions at $\th=0,\pm \pi$, the metric tensor $g_{\m\n}$
is a real, symmetric matrix, and therefore has $D(D+1)/2$
degrees of freedom at each spacetime point, modulo diffeomorphisms.
The metric can also be expressed in terms of vielbeins $g_{\m\n}=e^a_\m
\eta_{ab} e^b_\n$, and the vielbeins have $D^2$ degrees of freedom
(again, modulo diffeomorphisms).
Naturally, the number of gravitational degrees of freedom should be the
same, whether one counts metric or vielbein components.  In fact,
if $\eta_{ab}$ is the Euclidean or Minkowski metric, one should
subtract the dimension of the local Lorentz group ($O(D)$ or $O(D-1,1)$)
from the number of vielbein degrees of freedom, to get the actual
number of gravitational degrees of freedom. Since the
number of group generators is $D(D-1)/2$, we have for the inequivalent
vierbein degrees of freedom

\beq
      {D(D+1) \over 2} = D^2 - {D(D-1)\over 2}
\eeq
which is the same as the number of metric degrees of freedom.

   However, for the generalized metric, the "local Lorentz" invariance
is only \newline $O(D-1)$.
If the $e^a_\m$ are unrestricted, then the independent
vielbein degrees of freedom exceeds $D(D+1)/2$ except at $\th=0,\pm \pi$,
where the number is abruptly reduced.  Let us instead impose
\rf{con2}.  Then $e^0_\m$ contains only one degree of freedom, the
$e^i_\m~~(i\ne 0)$ contain $D(D-1)$ degrees of freedom, and subtracting
the dimension of the $O(D-1)$ group we have

\beq
      {D(D+1)\over 2} = 1 + D(D-1) - {(D-1)(D-2)\over 2}
\eeq
which is the usual number of gravitational degrees of freedom, modulo
diffeomorphisms.  Thus we can impose
\rf{con2} on the grounds that the dynamical Wick angle should not change
the number of independent degrees of freedom of the gravitational field.

   The final argument for the conditions \rf{con2} concerns fermionic
fields in curved space.  For the bosonic fields, the Lagrangian
involves the signature only via the metric, while for Dirac fields, the
signature also enters via the gamma matrices, which in the tangent
space should satisfy

\beq
          \{\g^a,\g^b\} = -2 \eta^{ab}
\label{gamma}
\eeq
The generalized Dirac action in curved space is just
the usual Dirac action
\beq
         S_D = \int d^D \sqrt{g} \overline{\psi} (-i\g^\m D_\m + m)\psi
\eeq
with
\bea
         \g^\m &=& e^\m_a \g^a
\non \\
          D_\m &=& \partial_\m + \oh \s^{ab}  \w_{\m ab}
\non\\
          \s^{ab} &=& \oq [\g^a,\g^b]
\non \\
          \w_{\m ab} &=&  e^\r_a e_{b\r;\m}
\label{Dirac}
\eea
where the $\g^a$ satisy \rf{gamma}.

  Equation \rf{Dirac} defines a spin-connection for covariant
derivatives acting on spinors at arbitrary $\th \in [-\pi,\pi]$.
The question is whether those covariant derivatives have the expected
properties.  Of course, since even global frame invariance is broken
at $\th \ne 0,\pm \pi$, we cannot demand that the spin-connection should
enforce local lorentz invariance for general $\th$.  Certain other
properties of the covariant derivative, however, are reasonable to
require.  Let us introduce a sort of "$ict$" notation
\bea
       g_{\m\n} &=& \et^a_\m \et^a_\n
\non \\
  \et^a_\m &=&    \left\{ \begin{array}{rr}
      e^{i\th/2} e^0_\m~~~~~~(a=0) \\
      e^a_\m~~~~~~~~~~~~(a \ne 0) \\
  \end{array} \right.
\non \\
      \{\g^a_E,\g^b_E\} &=& -2 \d^{ab}
\eea
where the latin indices of $\et$ and $\g_E$ are raised and lowered with the
Euclidean metric. In this notation, it is clear that the covariant derivative
should have the property

\bea
       0 &=& D_\m g_{\a\b} = D_\m (\et^a_\a \et^a_\b)
\non \\
         &=& (D_\m \et^a_\a)\et^a_\b + \et^a_\a(D_\m \et^a_\b)
\eea
which implies
\beq
      D_\m \et^a_\n = \et^a_{\n;\m} + \wt_{\m\;b}^{\;a} \et^b_\n = 0
\eeq
and therefore
\beq
      \wt_{\m ab} =  \et^\r_a \et_{\r b;\m}
\eeq
The covariant derivative for spinor fields in "$ict$" notation
would then be

\bea
       D^{"ict"}_\m &=& \partial_\m + \oh \s_E^{ab} \wt_{\m ab}
\non \\
        \s_E^{ab} &\equiv& \oq [\g^a_E,\g^b_E]
\eea
for arbitrary $\th$.  It then turns out that the covariant derivative
$D_\m$ in eq. \rf{Dirac} above, obtained simply by using generalized
Dirac gamma matrices in the Dirac action, and the covariant derivative
$D^{"ict"}_\m$ above, agree {\it only if conditions \rf{con2} are
imposed}.  In that case, it is easy to check that all derivatives
of $\th$ drop out of $\s^{ab} \w_{\m ab}$, in which case

\beq
      \s^{ab} \w_{\m ab} = \s^{ab}_E \wt_{\m ab}
\label{swt}
\eeq
and therefore $D_\m = D_\m^{"ict"}$.  A further consequence is that
$D_\m$ commutes with $\g^\m$, and by eq. \rf{gamma}, \rf{swt}, and
straightforward partial integration, one can verify that
\beq
      S_D = \int d^4x \; \overline{\psi} \left[i\overleftarrow{D_\m}
\g^\m + m \right] \psi \sqrt{g}
\eeq
leading to the standard equation for $\overline{\psi}$.
Similar considerations apply to the Weyl equation.

   The conclusion is that there are three separate reasons for
imposing the condition \rf{con2}, namely:

\begin{description}
\item[~~1.] ~To require that the metric
$g_{\m\n}=\eta_{\m\n}$ is flat;

\item[~~2.] ~To ensure that the number of gravitational degrees
of freedom (= inequivalent vielbein degrees of freedom)
is independent of the Wick angle;

\item[~~3.] ~To obtain a covariant derivative for spinors with
appropriate properties.
\end{description}

\noindent These conditions, taken together, also rule out having
more than one dynamical Wick angle in the tangent space metric,
as in eq. \rf{multiple}.  The reason is that requirements 1 and 3,
above, imply that $\partial_\m \th_a \propto e^a_\m$.  But then
the number of inequivalent vielbein degrees of freedom would be
{\it less} than $D(D+1)/2$, in violation of the second requirement.

    However, eq. \rf{con2} is a very severe restriction of $\th(x)$;
it means that rather than having one degree of freedom per point, which
is characteristic of a field, $\th(x)=\th(T(x))$ has only one degree of freedom
per $T=\mbox{const.}$ hypersurface, where the preferred time direction
$\partial_\m T$ is fixed by the choice of $e^0_\m$.  Obviously, a variable
which cannot vary locally is inimicable to the spirit of Lorentz invariance;
but local Lorentz invariance is lost, in any case, for any $\th\ne
0,\pm\pi$~.
\footnote{Diffeomorphism invariance, however, is an exact symmetry at
all $\th$.}  The whole argument of this paper is that Lorentz invariance
can arise dynamically; it does not have to be imposed from the beginning.

   We may now estimate the magnitude of fluctuations away from Lorentz
signature, in flat ($e^a_\m=\d^a_\m$) spacetime.  It is again assumed
that there is a high-frequency cutoff around the Planck scale, in
which case there is roughly one degree of freedom per
Planck-time. Writing $\th=\pi -\d \th$, the action for one Planck-time
(during which $\th$ is approximately constant) is

\bea
       \Delta S &\sim& \Lambda^4 V l_P \d \th
\non \\
                &\sim& {V \over l_P^3} \d \th
\eea
where $l_P$ is the Planck-length, and $V$ is the three-volume of the
$T=\mbox{const.}$ hypersurface.  Therefore
\beq
       <\d \th > \sim {l_P^3 \over V}
\eeq
Even under conservative assumptions, i.e. a closed Universe
of length scale on the order of $10^{10}$ light years, the ratio
of Planck-volume to the volume of the Universe gives
$\d \th \sim 10^{-184}$ radians.  It seems safe to say that
deviations from Lorentzian signature of this magnitude are
undetectable.  Of course, in the very early Universe, fluctuations
away from Lorentzian signature could have been substantial.

  Note that in this argument it was crucial that $\th$ is constant on
the preferred $T=\mbox{const.}$ hypersurfaces.  If this were not the case,
and $\th$ could vary locally, then entropy would overwhelm the
effective potential and we would instead expect $\d \th$ to be $O(1)$, which
is surely not consistent with observation.

\section{Cancellation of the Cosmological Term}

    The effective potential $V(\th)$ can be interpreted as a
$\th$-dependent cosmological "constant", and the argument of this
paper is based on looking for the minimum (of the real part)
and stationarity (for the imaginary part) of $V(\th)$ (the
"minimum/stationary point" of $V(\th)$).
Since $V(\pi)\ne 0$ in D=4 dimensions, the cosmological constant
is non-zero and of order $O(\L^4)$.  This, of course, raises the
question of how to justify expansion of the metric around
flat spacetime, in computing the one-loop
contribution to the determinant in eq. \rf{det}.

   It has been suggested occasionally that the cosmological term
is somehow \linebreak
screened at large distances \cite{Polyakov,Mottola},
and this  idea, if it really works, would justify the flat space expansion.
But it is obviously important to consider other possibilities.
The most conservative approach to the cosmological
constant problem is simply to add a counterterm

\beq
     S_c = \int d^Dx \; \sqrt{g} \lambda_c
\eeq
to remove the induced term.\footnote{The value of $\l_c$, like that
of all other bare masses and couplings, is assumed to be real.}
At first sight, it might seem that this
"conservative" approach  to removing the cosmological constant
also removes the mechanism which singles out Lorentzian signature
at D=4.  In fact, that is not true.  Writing
\beq
        \l = (n_F-n_B){\L^D \over D(4\pi)^{D/2}}
\eeq
the total effective potential is

\beq
      V_T(\th) = \l_c e^{i\th/2}+ \l e^{-i(D-2)\th/4}
\eeq
and it is clearly impossible to choose $\l_c$ such that $V_T(\th)=0$
for {\it all} $\th$.  Instead, the object would be to choose
$\l_c$ such that $V_T=0$ at the minimum/stationary point of
$V_T(\th)$.  It will now be shown that it is only possible
to make such a choice in $D=4$ dimensions, where the minimum/stationary
point is again Lorentzian signature.

    Denoting

\beq
      \thb \equiv {D-2 \over 2} \th
\eeq
the condition that $V_T=0$ gives

\beq
      \l_c \cos{\th \over 2} + \l \cos{\thb \over 2} = 0
\label{v1}
\eeq
for the real part,
\beq
      \l_c \sin{\th \over 2} - \l \sin{\thb \over 2} = 0
\label{v2}
\eeq
for the imaginary part, while
\beq
      \l_c \cos{\th \over 2} -{D-2\over 2} \l \cos{\thb \over 2} = 0
\label{v3}
\eeq
is the stationarity condition for $Im[V_T]$.  Equations \rf{v1} and \rf{v3}
imply that

\bea
              \thb &=& (2n+1)\pi
\non \\
              \th &=& \pi
\label{thb}
\eea
where we have used the fact that $|\th|\le \pi$ ($n$ integer).
Then, from \rf{v2} we have that
\beq
        |\l_c| = \l
\eeq
Now suppose $D$ is large enough so that one can choose $\thb \ge 3\pi$,
consistent with \rf{thb}.  The remaining question is whether $\th$
corresponding to this choice of $\thb$ is the mimimum point of
Re[$V_T$].  If $D$ is such that $\thb \ge 3\pi$ is possible, then
it would also be possible to choose a value of $\th=\th'$ where
$\thb'=(D-2)\th'/2=2\pi$, in which case

\bea
       \mbox{Re}[V_T(\th')] &=& \l_c \cos{\th'\over 2} - \l
\non \\
                   &=& -\l (1 \pm\cos{\th' \over 2})
\non \\
                   &<& 0
\eea
since $0<\th'\le \pi$. This would mean that $V_T=0$ is not
the minimum/stationary point, so the only other possiblity is
that $\thb=\pi$.  For $\th=\pi$, this can only be true in D=4
dimensions, in which case

\beq
        V_T(\th) = \l \cos{\th \over 2}
\eeq
and $\th=\pi$ is clearly the minimum of this potential.

\section{The case of massive fields}

   The analysis of the previous sections, applied to massless fields,
was extremely simple; it is not as simple when our considerations
are extended to massive fields.  The problem is
that the integral in eq. \rf{proper}, extended to massive fields,
involves incomplete gamma functions, and the corresponding analysis
becomes more involved.  Our approach will be to make an $m^2/\L^2$
expansion around $m=0$.  There are three cases of interest.
First of all, for $D=4$ and $n_F>n_B$, the mass corrections can be
expected to separate the minimum point (of $Re[V]$) and stationary
point (of $Im[V]$) slightly.  We will show below that this slight
separation does not destroy the Lorentzian behavior; it turns out that
the miminum of the real part is still exactly at $\th=\pm \pi$, while
the stationary point of the imaginary point moves just outside the
range $\th \in [-\pi,\pi]$, provided that a certain inequality among the
masses is satisfied.  Thus Lorentzian signature is still the optimum
$\th$ value. The other two cases of interest are $D=2$, and $n_F=n_B$.
For massless fields, $V(\th)$ is independent of $\th$ for those choices.
The introduction of masses can be expected to remove this degeneracy,
and the question is whether any new solutions of the
minimum/stationarity criteria are obtained.  We will find that only
for the case $n_B=n_F$ at $D=6$ is it possible to have the
minimum/stationary points (nearly) coincide.

  The starting point of our analysis is the one-loop
contribution $V_S(\th)$ to the effective potential due to the
integration over a scalar field $\phi$ of mass $m$ in a flat
background ($e^a_{\mu}=\delta^a_{\mu}$).  This is given by the obvious
extension of eq. \rf{det}, i.e.

\beq
\exp[-\int d^Dx V_S(\th)] = det^{-\oh}[- \sqrt{\eta}(\eta^{ab}
\partial_a \partial_b-m^2)]
\label{detm}
\eeq
Again evaluating the determinant with heat-kernel regularization,
one finds

\bea
V_S(\th)&=&-{1\over 2}\int^{\infty}_{1/\L^2} {ds\over s}
\int {d^Dp\over (2\pi)^D}\exp\left \{-s\left [e^{-i\th/2}p_0^2+
e^{i\th/2}(\vec{p}^{\;2}+m^2)\right ]\right \}
\non \\
&=&-{\exp[-i(D-2)\th/4]\over 2(4\pi)^{D/2}}\int^{\infty}_{1/\L^2} ds
{}~s^{-D/2-1}
\exp\left [- m^2e^{i\th/2}s\right ]
\label{properm}
\eea
The convergence of the $p$-integration still requires that $\th \in
[-\pi, \pi]$.

As in section 2, the one-loop contribution to $V_{eff}(\th)$ from each
bosonic (fermionic) propagating degree of freedom of mass $m_B$ ($m_F$)
turns out to be proportional to $det^{-1/2
(+1/2)}[-\sqrt{\eta}(\eta^{ab}\partial_a\partial_b-m^2_{B(F)})]$
(neglecting factors of $det^p(\eta)$, which one assumes to be absorbed
in the functional measure).  The heat-kernel regularized value of
these determinants is complicated, as compared to the massless
case of eq. \rf{proper}, by the exponential factor
$\exp\left [-m^2 e^{i\th/2}s \right ]$.
However, in the $p$-integration convergence domain, $\th \in [-\pi,
\pi]$, the integral in eq. \rf{properm} can be expressed in terms of the
incomplete gamma function \cite{bateman}.  Again $V(\th)$ is complex-valued
in general.  Summing over all one-loop bosonic (B) and fermionic (F)
contributions gives

\bea
Re[V(\th)] &=&{\L^D\over 4(4\pi)^{D/2}}\biggl \{\sum_F
x^{D/2}_F\left
[e^{i\th/2}\Gamma(-D/2,x_{F+})+e^{-i\th/2}\Gamma(-D/2,x_{F-})
\right ]
\non \\
&-&\sum_B x^{D/2}_B\left [e^{i\th/2}\Gamma(-D/2,x_{B+})+e^{-i\th/2}\Gamma
(-D/2,x_{B-})\right ]\biggr\}
\label{realm}
\eea

\bea
Im[V(\th)] &=&-{i\L^D\over 4(4\pi)^{D/2}}\biggl \{\sum_F
x^{D/2}_F\left
[e^{i\th/2}\Gamma(-D/2,x_{F+})-e^{-i\th/2}\Gamma(-D/2,x_{F-})
\right ]
\non \\
&-&\sum_B x^{D/2}_B\left [e^{i\th/2}\Gamma(-D/2,x_{B+})-e^{-i\th/2}\Gamma
(-D/2,x_{B-})\right ]\biggr\}
\label{imagm}
\eea
where we have defined

\beq
x_{B(F)\pm}\equiv {m^2_{B(F)}\over \L^2}~e^{\pm i\th/2}
\label{xdef}
\eeq
and $\Gamma(-D/2,x_{B(F)\pm})$ is the incomplete gamma function defined
in the Appendix (eqs. \rf{defgammauno} and \rf{defgammadue}).

The stationarity condition on $Im[V(\th)]$ is defined by

\beq
{\partial \over \partial\th}Im[V]\biggr\vert_{\thb}=0
\label{stat}
\eeq
Taking advantage of the useful relation \cite{bateman}

\beq
{\partial \over \partial\th}\Gamma(-D/2, x_{B(F)\pm})=\mp{i\over 2}
[x_{B(F)\pm}]^{-D/2}\exp [-x_{B(F)\pm}]
\eeq
eq. \rf{stat} becomes

\bea
0&=&\exp\left [i{(D-2)\over 4}\thb\right ]\biggl \{\sum_F
\left[(\overline\xmf)^{D/2}\gmf-e^{-\overline\xmf}\right]
\non \\
&-&\sum_B\biggl[(\overline\xmb)^{D/2}\gmb-e^{-\overline\xmb}\biggr]\biggr \}
\non \\
&+&\exp\left [-i{(D-2)\over 4}\thb \right ]\biggl \{\sum_F\biggl[
(\overline\xpf)^{D/2}\gpf -e^{-\overline\xpf}\biggr]
\non \\
&-&\sum_B\left[(\overline\xpb)^{-D/2}\gpb-e^{-\overline\xpb}\right]\biggr \}
\label{statm}
\eea
where a bar over the variable $x$ means that this has to be evaluated
at $\th=\thb$.

Eliminating the incomplete gamma function dependence of eq. \rf{realm}
by means of eq. \rf{statm}, one finds that the value of $Re[V(\th)]$
at the stationary point $\thb$ is

\bea
Re[V(\thb)]&=&\Lover\biggl
\{\sum_F\exp[-x_F\cos{\thb/2}]\cos{[(D-2)\thb/4+x_F\sin{\thb/2}]}
\non \\
&-&\sum_B\exp[-x_B\cos{\thb/2}]\cos{[(D-2)\thb/4+x_B\sin{\thb/2}]}\biggr \}
\label{realstat}
\eea
where $x_{B(F)}\equiv {m^2_{B(F)}\over \L^2}$, and we require, in
the stability range $\th \in [-\pi,\pi]$

\beq
\min{[Re[V(\th)]]} \approx Re[V(\thb)]
\label{minre}
\eeq
where the approximate equality of the minimum and stationary points is
up to $O(m^2/\L^2)$ corrections.  Typically, if masses are on the
order of the GUT scale and $\L$ is on the order of the Planck scale, we
would expect $m^2/\L^2 \approx 10^{-8}$.

We define:

\bea
\dnbf &\equiv &n_F-n_B
\non \\
\D (x^n)_{BF}&\equiv & \sum_F\left ({m_F^2\over \L^2}\right )^n-
\sum_B\left ({m_B^2\over \L^2}\right )^n
\non \\
\D (x^n\ln{x})_{BF}&\equiv & \sum_F\left ({m_F^2\over \L^2}\right )^n
\ln{\left ({m_F^2\over \L^2}\right )}-
\sum_B\left ({m_B^2\over \L^2}\right )^n
\ln{\left ({m_B^2\over \L^2}\right )}
\label{defdelta}
\eea
(where, now, $n_{B}$ and $n_F$ represent the total number -massless plus
massive- of bosonic and fermionic propagating degrees of freedom).
Since the GUT mass, on a logarithmic scale, is not so far from the
Planck mass, we will treat $\D (x^n)_{BF}$ and $\D (x^n\ln{x})_{BF}$
as being of the same order of magnitude.

The derivation of the expansions of eqs. \rf{realm},
\rf{imagm}, \rf{statm}  and \rf{realstat} is quite straightforward
although tedious, and the detailed $m^2/\L^2$ expansions for arbitrary
signature and dimension are collected in the Appendix.

   The three cases of interest, namely: (i) $n_F>n_B$ at $D=4$;
(ii) $D=2$; and (iii) $n_F=n_B$; will now be considered separately:

\paragraph{~~$\bullet$ ~~D=4 {\rm at} \bold ${\bf \dnbf >0}$}
\indent

   Except in the degenerate cases ($D=2$ or $n_F=n_B$), small mass
corrections cannot affect the conclusion of section 2 for $D \ne 4$,
namely, that the mimimum and stationary points are not close to one another.
For $D=4$ and $n_F>n_B$, however, mass terms will spoil the exact
coincidence of the two points.
In this case, only the second term of eq. \rf{statexpans} of the Appendix
trivially vanishes and the stationarity condition \rf{stat} becomes

\bea
0&\simeq &\dnbf\cos{\thb\over 2}+\left [\D (x^2\ln{x})_{BF}+{(2\g-1)
\over 2}\dxdue\right ]\cos{\thb\over 2}
\non \\
&-&{1\over 2}\D (x^2)_{BF}\thb\sin{\thb\over 2}
-{2\over 3}\D (x^3)_{BF}\cos{\thb}+O[\D (x^4\ln{x})_{BF}]
\label{quastanon}
\eea
whose approximate solution is

\beq
{\thb\over 2}\simeq \left ({2k+1\over 2}\right )\pi\left[1-{\D
(x^2)_{BF}\over\dnbf}\right ]+O[\D(x^3)_{BF}]
\label{quasolnon}
\eeq
($\gamma$ is the Euler constant, and $k=0, -1$).  For the real part of
$V(\th)$,
from eq. \rf{realexpans} of the Appendix evaluated at D=4, we find

\bea
Re[V(\th)]\vert_{D=4}&\simeq &{\L^4\over 2(4\pi)^2}\biggl
\{{1\over 2}\dnbf\cos{\th\over 2}-\dx
\non \\
&-&{1\over 2}\biggl [\dxduelog
+{(2\g-3)\over 2}\dxdue\biggr ]\cos{\th \over 2}
\non \\
&+&{1\over 4}\dxdue\th\sin{\th\over 2}
+O[\dxtre]\biggr \}
\label{quarenonth}
\eea

   In the stability range $\th \in [-\pi,\pi]$, the value $|\th|=\pi$
is still the mimimum of the real part of $V(\th)$ for $\dnbf >0$.
The stationary point
of the imaginary part is at $|\th| = \pi + \epsilon$, where $\epsilon$
is $O(\D (x^2))$.  Moreover, if

\beq
        \D (x^2)_{BF} < 0
\eeq
then $\epsilon > 0$, and the stationary point lies just outside the
stability range.  For masses at the GUT scale, and cutoff at the
Planck length, this means that the stationary point is

\beq
       |\thb| = \pi + O(10^{-16})
\eeq
which is certainly very close to the mimimum point at $|\th|=\pi$.  Moreover,
$\th=\pm \pi$ is as close as it is possible to come to the stationary
point in the stability range. We conclude that for $D=4$ at $n_F>n_B$,
Lorentzian signature is still the optimum value of $\th$, as in the massless
case.

\paragraph{~~$\bullet$ ~~D=2 {\rm at} \bold ${\bf\dnbf \neq 0}$}
\indent

  The approximate stationarity condition for $Im[V(\th)]$
in $D=2$ dimensions and arbitrary $\dnbf$ turns out to be

\bea
0&\simeq &[\D (x\ln{x})_{BF}+\g\dx]\cos{\thb\over 2}-{1\over 2}\D (x)_{BF}
\thb\sin{\thb\over 2}-\D (x^2)_{BF}\cos{\thb}
\non \\
&+&{1\over 4}\D (x^3)_{BF}\cos{3\thb \over 2}+O[\D (x^4\ln{x})_{BF}]
\label{duestanon}
\eea
while for the real part,

\bea
Re[V(\th)]\vert_{D=2}&\simeq &{\L^2\over 8\pi}\biggl \{\dnbf+
[\dxlog +(\g -1)\dx]\cos{\th \over 2}
\non \\
&-&{1\over 2}\dx\th\sin{\th\over 2}+O[\dxdue]\biggr \}
\label{duerenonth}
\eea
which follows from eq. \rf{realexpans} of the Appendix.

    Given that $\D (x)_{BF}$ is of the same order as $\D (x\ln{x})$,
and ruling out any special fine-tunings among the masses, there is no
reason at all that the stationary point of $Im(V)$ should coincide
with the minimum of $Re[V]$.

    Next we consider the $n_F=n_B$ case, separately in dimensions
D=2 through D=6 and $D>6$.

\paragraph{~~$\bullet$ ~~\bold ${\bf \dnbf =0}$ {\rm at} {\bf D=2}}
\indent

   The stationarity condition for $D=2$ with $\dnbf=0$ is identical
to the corresponding condition at $\dnbf \ne 0$, while the equation
for $Re[V]$ differs only by a constant.  Barring fine-tuning among
the masses, the minimum and stationary points are not close together.

\paragraph{~~$\bullet$ ~~\bold ${\bf \dnbf =0}$ {\rm at} {\bf D=3}}
\indent

In three dimensions, the stationarity condition becomes

\bea
0&\simeq &-2\D (x)_{BF}\cos{\thb\over 4}
+{8\over 3}\sqrt{\pi}\D (x^{3/2})_{BF}\cos{\thb\over 2}
-3\D (x^2)_{BF}\cos{3\thb \over 4}
\non \\
&+&{5\over 9}\dxtre \cos{5\thb\over 4}
+O[\D (x^4\ln{x})_{BF}]
\label{trestanon}
\eea
whose approximate solution is

\beq
{\thb\over 4}\simeq \left ({2k+1\over 2}\right )\pi+O\left [{\D
(x^{3/2})_{BF}\over \dx}\right ]
\label{tresol}
\eeq
This stationary point is well outside
the convergence domain $[-\pi, \pi]$.

\paragraph{~~$\bullet$ ~~\bold ${\bf \dnbf =0}$ {\rm at} {\bf D=4}}
\indent

   For the stationarity condition we have
\bea
0&\simeq & \left [\D (x^2\ln{x})_{BF}+{(2\g-1)
\over 2}\dxdue\right ]\cos{\thb\over 2}
\non \\
&-&{1\over 2}\D (x^2)_{BF}\thb\sin{\thb\over 2}
-{2\over 3}\D (x^3)_{BF}\cos{\thb}+O[\D (x^4\ln{x})_{BF}]
\eea
while for the real part

\bea
Re[V(\th)]\vert_{D=4}&\simeq &{\L^4\over 2(4\pi)^2}\biggl\{-\dx
+{1\over 4}\dxdue\th\sin{\th\over 2}
\non \\
&-&{1\over 2}\biggl [\dxduelog
+{(2\g-3)\over 2}\dxdue\biggr ]\cos{\th \over 2}
\non \\
&+&O[\dxtre]\biggr \}
\eea
and in general, the minimum/stationary points do not coincide.

\paragraph{~~$\bullet$ ~~\bold ${\bf \dnbf =0}$ {\rm at} {\bf D=5}}
\indent

In the case D=5, eq. \rf{statexpans} of the Appendix with $\dnbf=0$
becomes

\bea
0&\simeq &{2\over 3}\D (x)_{BF}\cos{\thb
\over
4}+\D (x^2)_{BF}\cos{\thb\over 4}-{16\over 15}\sqrt{\pi}\D (x^{5/2})_{BF}
\cos{\thb \over 2}
\non \\
&+&\dxtre\cos{3\thb\over 4}+O[\D (x^4\ln{x})_{BF}]
\label{cinstanon}
\eea
whose approximate solution is

\beq
{\thb\over 4}\simeq \left ({2k+1\over 2}\right )\pi +O\left [{\dxdue
\over \dx}\right ]
\label{cinsol}
\eeq
As in the previous case with D=3, the stationary point is far outside
the stability domain.

\paragraph{~~$\bullet$ ~~\bold ${\bf \dnbf =0}$ {\rm at} {\bf D=6}}
\indent

In six dimensions, eq. \rf{statexpans} of the Appendix
for the stationary point becomes

\bea
0&\simeq &\dx\cos{\thb\over 2}+{1\over 3}\left [\D
(x^3\ln{x})_{BF}+{(6\g-5)\over 6}\dxtre\right]\cos{\thb\over 2}
\non \\
&-&{1\over 6}\D (x^3)_{BF}\thb\sin{\thb\over 2}+O[\D (x^4\ln{x})_{BF}]
\label{seistanon}
\eea
whose solution is

\beq
{\thb\over 2}\simeq \left ({2k+1\over 2}\right )\pi\left[1-{1\over 3}{\D
(x^3)_{BF}\over\D (x)_{BF}}\right ]+O\left [{\Delta(x^4\ln{x})_{BF}\over
\dx} \right ]
\label{seisol}
\eeq
Therefore, the stationary point of $Im[V(\th)]$ can be just
outside $[-\pi, \pi]$ if the following inequality holds

\beq
{\dxtre\over \dx}<0
\label{seisolth}
\eeq
Moreover, eq. \rf{realexpans} of the Appendix gives

\beq
Re[V(\th)]\vert_{D=6}\simeq {\L^6\over 2(4\pi)^3}\biggl \{
-{1\over 2}\dx\cos{\th\over 2}
+O[\dxtrelog]\biggr \}
\label{seirenonth}
\eeq
and this has a minimum in the stability domain exactly at $\th=\pm \pi$,
if
\beq
\dx <0
\label{seisolre}
\eeq

   This means that the case of $D=6$ and $\dnbf=0$ is similar to
$D=4$ and $n_F>n_B$.  Assuming two inequalities, namely $\dx<0$ and
$\dxtre >0$, we find that $\mbox{minRe}$[$V]$ is at $|\th|=\pi$, and
the stationary
point of $\mbox{Im}$[$V]$ is at $|\th|=\pi+\epsilon$, where $\epsilon$ is
positive and $O(m^4/\L^4)$.  As in the D=4 case, Lorentzian signature
is the optimum $\th$ value in the range $[-\pi,\pi]$.

\paragraph{~~$\bullet$ ~~\bold ${\bf \dnbf =0}$ {\rm at} {\bf D}\bold
${\bf >}${\bf 6}}
\indent

Finally, we consider the cases D$>$6 with $\dnbf=0$.
For the stationary part

\bea
0&\simeq &2\left ({D-4\over D-2}\right )\dx\cos\left[{(D-4)\thb\over 4}\right ]
-\left ({D-6\over D-4}\right )\dxdue\cos\left[{(D-6)\thb\over 4}\right ]
\non \\
&+&{1\over (D-6)}\biggl[{D-8\over 3}
+{32(1-h(D-8))\over D(D-2)(D-4)}\cdp\biggr ]
\dxtre\cos\left[{(D-8)\thb\over 4}\right ]
\non \\
&+&O[\D (x^4\ln{x})_{BF}]
\label{setstanon}
\eea
so that

\beq
{(D-4)\over 4}\thb\simeq \left ({2k+1\over 2}\right )\pi
+O\left[{\dxdue\over \dx}\right]
\label{setsol}
\eeq
while for the real part

\beq
Re[V(\th)]\vert_{D>6}\simeq {\L^D\over 2(4\pi)^{D/2}}\biggl \{
-{2\over D-2}\dx\cos\left[{(D-4)\th\over 4}\right ]
+O[\dxdue]\biggr \}
\label{setrenonth}
\eeq
It is readily seen from eq. \rf{setrenonth} that,
in general, the minimum of $Re[V(\th)]$ is not at $\thb$.
This eliminates from consideration all dimensions $D>6$.

  We have, throughout, treated $\dx$ and $\dxlog$ as being of the
same order of magnitude.  If the Planck scale is {\it not} a fundamental
cutoff, so that $\L$ can be taken arbitrarily large, or if the mass
generation scale is many orders of magnitude
less than the presumed grand unification scale, then it is appropriate
to treat $\dxlog >> \dx$.  In that case, in addition to Lorentzian
solutions at $D=4~~(n_F>n_B)$ and $D=6~~(n_F=n_B)$ we find additional
Lorentzian
solutions at $D=2~~(\dnbf~\mbox{arbitrary})$, and $D=4~~(n_F=n_B)$.

  Finally, since non-zero mass terms displace the stationary point
slightly away from the minimum point at $D=4~~(n_F>n_B)$ and
$D=6~~(n_F=n_B)$, the exact cancellation of the cosmological constant
found in section 4 is no longer quite exact.  Although the real part of
$V_T(\th)$ can be cancelled exactly at the minimum point ($|\th|=\pi$),
one would expect a small imaginary part, of order $m^4/\L^4$,
left over, which in principle constitutes a contribution to the measure.

\section{Conclusions}

   Two fundamental facts about spacetime are its Lorentzian
signature and D=4 dimensionality.  An equally fundamental
feature of quantum mechanics, which distinguishes it from any
sort of classical field theory or diffusion process, is the appearance
of $\sqrt{-1}$ in the Feynman amplitude and Schrodinger equation.
The proposal that spacetime signature (i.e. the tangent
space metric) is dynamical provides an intriguing relation among these three
facts.  The $i$ of quantum mechanics can be traced to the factor
$\sqrt{g}=\vert e\vert\sqrt{\eta}$ in the path amplitude, which becomes just
$\exp[iS]$ at Lorentzian signature.  By allowing the tangent-space
metric $\eta_{ab}$ to interpolate {\it continuously} between different
signatures (which requires that entries of $\eta$ can rotate into
the complex plane), we have found by a simple one-loop argument that
Lorentzian signature is dynamically selected, for $n_F>n_B$, uniquely
in $D=4$ dimensions.  In broken supersymmetry theories,
there is also a possibility for Lorentzian signature at $D=6$.
With the help of curved-space consistency
conditions, it has been further argued that fluctuations away from
Lorentzian signature at $D=4$ are enormously suppressed (except, perhaps,
in the very early Universe) and are certainly undetectable in the
present epoch.


\newpage

\noindent {\Large \bf Acknowledgements}{\vspace{11pt}}

  The hospitality of the Lawrence Berkeley Laboratory,
where much of this work was carried out, and the support of the
Danish Research Council, is gratefully aknowledged.
J.G. also thanks Ron Adler and Alexander Yelkhovsky for helpful
discussions.

\vspace{33pt}

\appendix

\section{Appendix}
In this Appendix we give a list of basic definitions and equations,
which we use and to which we make reference in the main text,
for the case of arbitrary signature and dimensions.

In general, the incomplete gamma function $\Gamma(\alpha, x)$ is
described by two different series expansions in the variable $x$ for
the case of $\alpha$ integer (D even) or fractional (D odd).
The two expansions can be combined by writing

\bea
\Gamma(-D/2, x_{\pm})&=&{(-1)^{D/2}\over \left ({D\over 2}\right )!}
\left [E_1(x_{\pm})-e^{-x_{\pm}}\sum_{n=0}^{D/2-1}{(-1)^nn!\over
(x_{\pm})^{n+1}}\right ]\cdp
\non \\
&+&\left [{(-1)^{(D+1)/2}\sqrt{\pi}\over \left ({D\over
2}\right )!}-\sum_{n=0}^{\infty}{(-1)^n
(x_{\pm})^{n-D/2}\over
n!(n-D/2)}\right ]\sdp
\label{defgammauno}
\eea
where $E_1(x_{\pm})$ is the exponential integral
function

\beq
E_1(x_{\pm})\equiv -\left
(\g+\ln{x_{\pm}}+\sum_{n=1}^{\infty}{(-x_{\pm})^n\over n!n}\right )
\label{defgammadue}
\eeq
and $\g$ is the Euler constant \cite{bateman}.
The effect of $\cdp$ ($\sdp$) in eq. \rf{defgammauno} is just to select
out one of the two (exact) expansions for the case when D is even
(odd) \cite{bateman}.

Moreover, using the series expansion \rf{defgammauno} and assuming
$x\ll 1$, one can easily rewrite the stationarity condition for
$Im[V]$,
eq. \rf{statm} of section 5, up to order $\ox$ as\footnote{
We restrict our analysis to the case of dimension D$\geq 2$, i.e. we
do not consider the case of a D=1, single particle quantum mechanics.}
\bea
0&\simeq &{2(2-D)\over D}\dnbf\cos{\left[{(D-2)\thb\over
4}\right]}+2\biggl[{(D-4)\over (D-2)}\sdp
\non \\
&+&{1\over D}\left (D-2-{4h(D-4)\over (D-2)}\right )
\cos^2{\left ({D\pi\over 2}\right)}\biggr ]\D (x)_{BF}\cos{\left[{(D-4)
\thb\over 4}\right]}
\non \\
&+&(-1)^{D/2+1}{2\over \left({D\over 2}\right)!}\D (x^{D/2}\ln{x})_{BF}\cdp
\cos{\left ({\thb\over 2}\right )}h(6-D)
\non \\
&+&(-1)^{D/2+1}{2\over \left({D\over 2}\right)!}\D (x^{D/2})_{BF}\biggl [
\biggl (\g\cdp
\non \\
&+&(-1)^{-1/2}\sqrt{\pi}\sdp\biggr )\cos{\left ({\thb\over 2}\right )}
-{\thb\over 2}\sin{\left ({\thb\over 2}\right )}\cdp \biggr]h(6-D)
\non \\
&+&(-1)^{D/2}{2\over \left({D\over 2}\right)!}\D
(x^{D/2+1})_{BF}\cdp\cos{(\thb)}h(4-D)
\non \\
&+&{(-1)^{D/2+1}\over 2\left({D\over 2}\right)!}\D (x^{D/2+2})_{BF}\cdp\cos
{\left ({3\thb\over 2}\right)}h(2-D)
\non \\
&-&\biggl[{(D-6)\over (D-4)}\sdp+{1\over D}\biggl(D-2
-{16h(D-6)\over (D-2)(D-4)}
\non \\
&-&{8h(D-4)\over (D-2)}\biggr)
\cdp\biggr]\D (x^2)_{BF}\cos{\left[{(D-6)\thb\over 4}\right]}
\non \\
&+&\biggl[{(D-8)\over 3(D-6)}\sdp+{1\over D}\biggl({D-2\over 3}
-{32h(D-8)\over (D-2)(D-4)(D-6)}
\non \\
&-&{16h(D-6)\over (D-2)(D-4)}-{4h(D-4)\over (D-2)}\biggr)
\cdp\biggr]\D (x^3)_{BF}\cos{\left[{(D-8)\thb\over
4}\right]}
\non \\
&+&O[\D (x^4\ln{x})_{BF}]
\label{statexpans}
\eea
where $h(y)$ is the Heaviside step function

\bea
h(y)&\equiv & \left\{ \begin{array}{ll}
1 &\mbox{$y\geq 0$}\\
0 &\mbox{$y<0$}\\
\end{array}\right.
\label{heavy}
\eea

Similarly, expansion of eqs. \rf{realm} and \rf{imagm} of section 5 gives

\bea
Re[V(\th)]&\simeq &\Lover\biggl \{{2\over D}\dnbf\cos{\left[{(D-2)\th
\over 4}\right]}-2\biggl[{1\over (D-2)}\sdp
\non \\
&+&{1\over D}\left (1+{2h(D-4)\over (D-2)}\right )
\cos^2{\left({D\pi\over 2}\right)}\biggr ]\D (x)_{BF}\cos{\left[{(D-4)
\th\over 4}\right]}
\non \\
&+&{(-1)^{D/2+1}\over \left({D\over 2}\right)!}\D (x^{D/2}\ln{x})_{BF}\cdp
\cos{\left ({\th\over 2}\right )}h(6-D)
\non \\
&+&{(-1)^{D/2+1}\over \left({D\over 2}\right)!}\D (x^{D/2})_{BF}\biggl[
\biggl(\g\cdp
\non \\
&+&(-1)^{-1/2}\sqrt{\pi}\sdp\biggr)\cos{\left({\th\over 2}\right)}
-{\th\over 2}\sin{\left({\th\over 2}\right)}\cdp\biggr]h(6-D)
\non \\
&+&{(-1)^{D/2}\over \left({D\over 2}\right)!}\D (x^{D/2+1})_{BF}\cdp
\cos{(\th)}h(4-D)
\non \\
&+&{(-1)^{D/2+1}\over 4\left({D\over 2}\right)!}\D (x^{D/2+2})_{BF}\cdp
\cos{\left({3\th\over 2}\right)}h(2-D)
\non \\
&+&\biggl[{1\over (D-4)}\sdp+{1\over D}\biggl(
1+{8h(D-6)\over (D-2)(D-4)}
\non \\
&+&{4h(D-4)\over (D-2)}\biggr)
\cdp\biggr]\D (x^2)_{BF}\cos{\left[{(D-6)\th\over 4}\right]}
\non \\
&+&\biggl[-{1\over 3(D-6)}\sdp-{1\over D}\biggl(
{1\over 3}+{16h(D-8)\over (D-2)(D-4)(D-6)}
\non \\
&+&{8h(D-6)\over
(D-2)(D-4)}+{2h(D-4)\over (D-2)}\biggr)
\cdp\biggr]\D (x^3)_{BF}\cos{\left[{(D-8)\th\over
4}\right]}
\non \\
&+&O[\D (x^4\ln{x})_{BF}]\biggr \}
\label{realexpans}
\eea

\bea
Im[V(\th)]&\simeq &\Lover\biggl \{-{2\over D}\dnbf\sin{\left[{(D-2)\th
\over 4}\right]}+2\biggl[{1\over (D-2)}\sdp
\non \\
&+&{1\over D}\left (1+{2h(D-4)\over (D-2)}\right )
\cos^2{\left({D\pi\over 2}\right)}\biggr]\D (x)_{BF}\sin{\left[{(D-4)
\th\over 4}\right]}
\non \\
&+&{(-1)^{D/2+1}\over \left({D\over 2}\right)!}\D (x^{D/2}\ln{x})_{BF}\cdp
\sin{\left ({\th\over 2}\right )}h(6-D)
\non \\
&+&{(-1)^{D/2+1}\over \left({D\over 2}\right)!}\D (x^{D/2})_{BF}\biggl[
\biggl(\g\cdp
\non \\
&+&(-1)^{-1/2}\sqrt{\pi}\sdp\biggr)\sin{\left({\th\over 2}\right)}
+{\th\over 2}\cos{\left({\th\over 2}\right)}\cdp\biggr]h(6-D)
\non \\
&+&{(-1)^{D/2}\over \left({D\over 2}\right)!}\D (x^{D/2+1})_{BF}\cdp
\sin{(\th)}h(4-D)
\non \\
&+&{(-1)^{D/2+1}\over 4\left({D\over 2}\right)!}\D (x^{D/2+2})_{BF}\cdp
\sin{\left({3\th\over 2}\right)}h(2-D)
\non \\
&-&\biggl[{1\over (D-4)}\sdp+{1\over D}\biggl(
1+{8h(D-6)\over (D-2)(D-4)}
\non \\
&+&{4h(D-4)\over (D-2)}\biggr)
\cdp\biggr]\D (x^2)_{BF}\sin{\left[{(D-6)\th\over 4}\right]}
\non \\
&+&\biggl[{1\over 3(D-6)}\sdp+{1\over D}\biggl(
{1\over 3}+{16h(D-8)\over (D-2)(D-4)(D-6)}
\non \\
&+&{8h(D-6)\over
(D-2)(D-4)}+{2h(D-4)\over (D-2)}\biggr)
\cdp\biggr]\D (x^3)_{BF}\sin{\left[{(D-8)\th\over
4}\right]}
\non \\
&+&O[\D (x^4\ln{x})_{BF}]\biggr \}
\label{imagexpans}
\eea

and, finally, eq. \rf{realstat} of section 5 becomes

\bea
Re[V(\thb)]&\simeq &\Lover\biggl \{\cos{\left[{(D-2)\thb\over 4}\right]}
\biggl[\dnbf-\D (x)_{BF}\cos{\thb\over 2}
\non \\
&+&{1\over 2}\D (x^2)_{BF}\cos{\thb}-
{1\over 6}\D (x^3)_{BF}\cos{\thb\over 2}\left (4\cos^2{\left({\thb\over
2}\right)}-3\right)\biggr]
\non \\
&-&\sin{\left[{(D-2)\thb\over 4}\right]}
\biggl[\D (x)_{BF}\sin{\thb\over 2}-{1\over 2}\D (x^2)_{BF}\sin{\thb}
\non \\
&-&{1\over 6}\D (x^3)_{BF}\sin{\thb\over 2}\left (4\sin^2{\left({\thb\over
2}\right)}-3\right)\biggr]+O[\Delta(x^4)_{BF}]\biggr \}
\label{realstatexpans}
\eea

\vfill\eject


\end{document}